\documentclass[12pt]{article}

\usepackage{newtxtext,newtxmath}

\usepackage{graphicx}

\usepackage[letterpaper,margin=1in]{geometry}

\renewenvironment{abstract}
	{\quotation}
	{\endquotation}

\date{}

\makeatletter
\renewcommand{\fnum@figure}{\textbf{Figure \thefigure}}
\renewcommand{\fnum@table}{\textbf{Table \thetable}}
\makeatother

\usepackage{scicite}

\usepackage{url}

\usepackage{float} 

\def\scititle{
	Leveraging target dynamics for imaging in complex media
}
\title{\bfseries \boldmath \scititle}

\author{
    Yoav~Ben~Haim$^{1\dagger}$,
	Ilay~Tomer$^{1\dagger}$,
	Omri~Haim$^{1}$,
	Benzy~Laufer$^{1}$,
	Ori~Katz$^{1\ast}$\and
    \small$^{1}$Institute of Applied Physics, The Hebrew University of Jerusalem, Jerusalem 9190401, Israel.\and
    \small$^\ast$Corresponding author. Email: orik@mail.huji.ac.il\and
	\small$^\dagger$These authors contributed equally to this work.
}

\begin{document} 

\maketitle

\begin{abstract} \bfseries \boldmath
Optical imaging in complex samples such as biological tissues is fundamentally challenging due to random light scattering that degrades resolution and contrast. When imaging realistic targets that contain natural dynamics such as flowing blood, the temporal variability introduces an additional obstacle, as the leading computational scattering-compensation methods require the target to remain stationary during a multi-frame acquisition process. Here we show that instead of struggling to perform rapid acquisitions, the target dynamics themselves can serve as an intrinsic information source for scattering compensation, replacing multiple controlled illuminations. The mathematical equivalence between target dynamics and conventional varying illumination patterns allows to demonstrate this approach in coherent holographic imaging and incoherent fluorescence microscopy using established matrix and model-based scattering-compensation techniques. Our general framework enables reconstruction of dynamic scenes with a number of acquisitions equal to the number of reconstructed frames, without the use of any spatial light modulators or illumination control.

\end{abstract}

\noindent
A common scenario in imaging through complex scattering or aberrating media, such as biological tissue, contaminated or damaged optical elements, or fog, is that light scattering drastically degrades resolution and image contrast \cite{bertolotti2022imaging}. 
In addition, for naturally dynamic targets, such as live tissue containing flowing and perfusing blood, the temporal variability of the dynamic scene introduces an additional challenge, limiting the acquisition time to the dynamics timescales, and requiring the scattering compensation scheme to take into account \textit{temporal} variations in the sample.

Recently developed approaches to mitigate the effects of severe scattering have shown great advancements in static scenarios:  wavefront shaping \cite{vellekoop2007focusing, gigan2022roadmap,bertolotti2022imaging}, where a computer-controlled spatial light modulator (SLM) phase pattern is optimized to undo scattering, has been demonstrated to allow noninvasive imaging  \cite{yeminy2021guidestar}, including fluorescence microscopy \cite{Aizik2024,monin2025, feng23}.  However, these techniques require a relatively large set of measurements under different wavefront-shaped patterns to determine the required SLM correction mask(s), which is challenging when imaging dynamic scenes, in particular with low photon-flux.

A different set of approaches for image reconstruction is derived from speckle correlation techniques, first introduced for astronomical observations  \cite{labeyrie1970attainment, ayers1988knox}. These computational-reconstruction techniques exploit the optical "memory effect" \cite{freund1988memory}, i.e., the shift-invariance (or isoplanatism) of the scattering point spread function (PSF), to retrieve target images from a single captured speckle image \cite{katz2014non, wu2016single}. Yet, these computational approaches are limited to relatively simple and sparse targets, require capturing a very large number of speckle grains in a single shot, and may rely on phase-retrieval algorithms that are sensitive to noise and lack robust convergence for complex targets \cite{katz2014non,fienup1978reconstruction}. 

A more powerful set of computational approaches is based on reflection-matrix processing \cite{Kang2017,badon2020distortion, Farah2024}. In these schemes, the scattering response of the sample to different input illuminations is measured, stored in a matricial form, and computationally decomposed into the target object image and the scattering components \cite{Kang2017, Kang2025}.
Matrix-based techniques require hundreds to thousands of measurements of scattered light under varying illumination conditions, which can be realized by a set of plane waves \cite{Kang2017}, scanned focused beams \cite{badon2020distortion}, or dynamic random speckle illumination \cite{lee22,weinberg2024fluorescence}. 
However, the conventional application of matrix-based methods works under the assumption that both the targets and the scattering (or aberrations) remain stationary throughout the multi-frame acquisition process, making imaging of dynamic scenes extremely challenging.

Recently, model-based computational wavefront shaping schemes where the scattering is digitally corrected without the need for an SLM were demonstrated to allow computational scattering correction using a considerably smaller number of captured frames than matricial approaches \cite{Haim2025, Zhang2025Adaptive}. 
These methods find the scattering correction by gradient-descent optimization of the wavefront correction to maximize the image quality.
However, these methods generally also require that the object remain static during the multi-frame acquisition. 
Neural-network model-based computational approaches \cite{feng23, xie2024wavemo, Jiang2026} can handle dynamics by incorporating temporal changes into the neural network representation, but have currently been applied only to relatively simple motions and aberrations. 

Here, we show that random uncontrolled dynamics of natural targets such as flowing blood can, in fact, serve as a source of information for computational scattering compensation rather than being an obstacle to imaging through scattering media. Specifically, we demonstrate that natural dynamic temporal variations, including \textit{in-vivo} blood flow, can be mathematically equivalent to the varying illumination used in conventional reflection-matrix measurements \cite{lee22} or model-based computational wavefront shaping \cite{Haim2025}.
We present proof-of-principle experiments in both coherent holographic imaging and incoherent fluorescence imaging through scattering layers. We show that various computational approaches, including matrix-based compensation \cite{weinberg2024fluorescence} and image-guided model-based correction \cite{Haim2025}, can successfully reconstruct the dynamic targets using the target dynamics themselves as a source of information.

The presented framework enables the recovery of both the dynamic targets and the scattering from a simple set of scattered light frames that are captured by a conventional widefield microscope while the target is dynamically varying. This extends the applicability of the state-of-the-art computational scattering compensation schemes to dynamic scenarios.

\section*{RESULTS}
\subsection*{Principle}

\noindent\
Figure~\ref{fig1_principle} outlines the concept of our computational approach for imaging dynamic targets through a static scattering medium. The concept applies to both coherent holographic imaging and incoherent imaging. For simplicity, we present below the derivation for the case of coherent holographic imaging. A similar derivation holds for the case of incoherent (e.g., fluorescence) imaging and is provided in the Supplementary Text S1. 
In coherent holographic imaging of a dynamic target through static scattering media (Fig.~\ref{fig1_principle}B), the measured scattered field $E_\text{m}\!\left(\vec{r}\right)$ at time $t_\text{m}$ can be expressed as: 
\begin{equation}\label{measure field integral}
E_\text{m}\!\left(\vec{r}\right) = \!\int\! P\!\left(\vec{r}, \vec{r}\,'\right) 
\,\left[O_\text{m}\!\left(\vec{r}\,'\right) 
\, E^\text{ill}\!\left(\vec{r}\,'\right) \right]
\, d\vec{r}\,'
\end{equation}
\noindent
where $P\!\left(\vec{r}, \vec{r}\,'\right)$ is the static scattering coherent point spread function (PSF) that describes the propagation of light from point $\vec{r}\,'$ at the object plane,  through the scattering medium, to point $\vec{r}$ at the detector plane, $O_\text{m}\!\left(\vec{r}\,'\right)$ represents the temporally-varying complex reflectance of the target at time $t_\text{m}$, and $E^\text{ill}\!\left(\vec{r}\,'\right)$ is the temporally fixed illumination at the target plane. Any temporal variations in the illumination can be effectively embedded into the temporally varying object function, $O_\text{m}\!\left(\vec{r}\,'\right)$.
Notably, Eq.~\ref{measure field integral} which describes the measured fields in the case of dynamic targets, has a similar form to the equation that describes the measured fields in conventional reflection matrix-based acquisition of static targets using randomly varying illumination\cite{lee22}, $E_\text{m}^\text{ill}\!\left(\vec{r}\,'\right)$ (Fig.~\ref{fig1_principle}A):
\begin{equation}
E_\text{m}\!\left(\vec{r}\right) = \!\int\! P\!\left(\vec{r}, \vec{r}\,'\right) 
\, \left[O\!\left(\vec{r}\,'\right) 
\, E_\text{m}^\text{ill}\!\left(\vec{r}\,'\right) \right]
\, d\vec{r}\,'
\end{equation}

The two cases are indeed mathematically equivalent as long as the target average reflected intensity pattern throughout the acquisitions, $\langle|O_m(\vec{r})|^2\rangle_m$, has some spatial features, i.e., the averaged reflected intensity is not completely homogeneously distributed across the field of view (FoV). This is required to allow image-metric-based optimization \cite{Haim2025} or matrix decomposition algorithms such as CTR-CLASS (compressed time reversal closed-loop accumulation of single scattering)\cite{lee22} that rely on the target having non-zero k-space components. 
An additional requirement is that the temporal dynamics between different spatial positions on the target are uncorrelated, since in matrix-based correction a virtual reflection matrix is constructed from the covariance matrix of the captured frames \cite{weinberg2024fluorescence} (see Supplementary Text S2), and for image-guided model-based correction an effective spatially-incoherent illumination is required \cite{Haim2025}. In the case where a fixed background or slowly varying global changes induce undesired correlations between different spatial positions across the target, one can use the differences between consecutive frames as the input to the computational correction algorithms, as we demonstrate below (see Supplementary Text S4).
With the above conditions met, scattered fields that are captured while such targets are dynamically varying can be processed with conventional matrix decomposition scattering compensation algorithms \cite{lee22, lee2023exploiting,badon2020distortion, balondrade2024, kang2023tracing, weinberg2024fluorescence, gupta2025imaging} or with computational model-based wavefront shaping  \cite{Haim2025}, to reconstruct the hidden dynamic target images and the scattering PSF from a set of rapidly-acquired conventional widefield frames.

Thus, our reconstruction process for the case of coherent imaging follows the same framework as CTR-CLASS  \cite{lee22}, or computational wavefront-shaping \cite{Haim2025}, and follows the framework of I-CLASS (incoherent CLASS) \cite{weinberg2024fluorescence} for the incoherent imaging cases. 
In short, for the matrix-based frameworks (see Supplementary Text S2), we use the covariance matrix of the measured datasets, i.e., the cross-correlations between any two pixels in the captured frames across the random temporal realizations. This covariance matrix has the same form as a "virtual reflection matrix" \cite{lee22, weinberg2024fluorescence}  where each pixel serves both as a detector and a virtual source, akin to passive correlation imaging using ambient noise in seismology \cite{Snieder2010}, and is then processed using I-CLASS variant of the established CTR-CLASS algorithm \cite{weinberg2024fluorescence}. 
For the computational wavefront-shaping framework,  we directly apply the model-based gradient descent optimization algorithm of Haim et al. \cite{Haim2025} to the captured dataset. This algorithm searches for a digital "virtual SLM" correction phase-pattern that would provide the sharpest corrected target image, as defined by an image quality metric such as the contrast of the incoherently compounded reconstructed image intensity pattern (see Supplementary Text S3).
In experiments where the scattering is anisoplanatic, i.e., the scattering PSF varies across the FoV, we divided the FoV into smaller isoplanatic patches, corrected each one independently, and then stitched the corrected patches together (see Supplementary Text S5).

\subsection*{Experimental results: holographic coherent imaging}

As a first controlled demonstration (Fig.~\ref{fig2_tube}), we imaged optically absorbing red Polystyrene microspheres, of 6~$\mu$m diameter, flowing through a transparent tube, mimicking blood flow in vessels. Figure~\ref{fig2_tube}A shows the off-axis holography setup used to record the scattered fields. The polystyrene microspheres in water suspension (for details see Materials and Methods) flow through a silicone tube with an inner diameter of 0.3~mm and an external diameter of 0.8~mm, which was immersed in water. A 632.8~nm Helium–Neon laser illuminates the flowing beads through a static diffuser ($0.5^\circ$) placed $ 53~mm$  from the target, the back-reflected light scatters through the same diffuser and is holographically recorded by a CCD camera imaging the scattering diffuser plane. A total of $M=180$ short-exposure frames (5\~ms exposure time)   are captured at a frame-rate of 17~fps.
 
As illustrated in Fig.~\ref{fig2_tube}B, direct Fresnel back-propagation (refocusing) to the object plane results in a set of low-resolution blurred images. Similarly, fluctuation imaging of the complex fields \cite{Chaigne2017}
(temporal standard deviation of the dynamic measure, $\sigma(r) = \sqrt{\frac{1}{M} \sum_{m=1}^{M} \left|E_m(r) - \langle E(r) \rangle_m\right|^2}$, where $\langle E(r) \rangle_m = \frac{1}{M}\sum_{m=1}^{M} E_m(r)$) without scattering, which is analogous to dynamic speckle illumination \cite{ventalon2005quasi}, fails to resolve the tube boundaries.

In this experiment, the static reflectivity of the tube walls exhibited a slow phase drift, which created temporal correlations between different portions of the target. These violate the requirement for uncorrelated fluctuations. To address this practically common issue we applied a differential-imaging step where we subtracted consecutive frames before further processing. This acts as a temporal high-pass filter, isolating the desired uncorrelated fast-varying signals from the slowly varying background, which had timescales of $>10$ frames. A simple mean subtraction, the standard alternative for background removal, was ineffective due to the slowly varying phase drifts. 

Following this simple preprocessing, we corrected the anisoplanatic scattering using a mosaicking approach (Supplementary Text S5). Since the scattering PSF varied across the FoV, the image was divided into nine approximately isoplanatic patches; each was independently reconstructed and subsequently stitched together. Figures~\ref{fig2_tube}D-F and \ref{fig2_tube}G-I present side-by-side comparisons of matrix-based (memory-efficient implementation of CTR-CLASS \cite{weinberg2024fluorescence}) and model-based computational wavefront-shaping correction \cite{Haim2025}, respectively, yielding similar results for this simple scenario. Compared to the raw field back-propagated to the object plane (Fig.~\ref{fig2_tube}B), both methods successfully resolve the tube structure and the flowing beads (Figs.~\ref{fig2_tube}D,G; Supplementary Movie S1).

Our second example (Fig.~\ref{fig3_finger}) uses the same experimental setup to demonstrate imaging using the intrinsic fast blood-flow dynamics in a human finger. Although a human finger appears static to the naked eye under incoherent illumination, substantial temporal fluctuations at sub-millisecond timescales from blood flow and perfusion are present \cite{Qureshi2017}. 
A USAF resolution target was placed in front of the finger as the target object to be imaged (Fig.~\ref{fig3_finger}).

As shown in Fig.~\ref{fig3_finger}B, any single captured field that is digitally back-propagated to the object plane without scattering correction exhibits severe scattering distortions. This is also true for the temporal fluctuation image computed from all propagated fields (Fig.~\ref{fig3_finger}C). However, applying the CTR-CLASS algorithm on the acquired dataset reveals the fine features of the target. Since the target extended beyond a single isoplanatic patch, we employed here as well a mosaicking approach (see details in Materials and Methods and in Supplementary Text~S5).

The temporal standard deviation of the corrected fields is presented in Fig.~\ref{fig3_finger}D, where the different reconstruction patches are color-coded for visualization. The corresponding recovered phase masks for the different patches are shown in Fig.~\ref{fig3_finger}E, revealing spatially varying aberrations across the FoV. Notably, the reconstruction relied entirely on the naturally rapid blood-flow induced intrinsic temporal dynamics, without external modulation or illumination control.

\subsection*{Experimental results: incoherent fluorescence microscopy}

To demonstrate our approach for incoherent imaging, we applied it on a dataset acquired in a conventional widefield fluorescence microscope setup, imaging flowing fluorescent beads behind a static scattering layer in an epi-fluorescence configuration. The optical setup is shown in Fig.~\ref{fig4_incoherent}A.  The sample consisted of 10~$\mu$m-diameter YG fluorescent beads flowing inside a transparent tube, placed inside a water-filled cuvette
(see Materials and Methods). Relatively uniform illumination was provided by a continuous-wave 488~nm laser, homogenized by passing through a rapidly rotating diffuser in the illumination path (not shown). The scattered fluorescence emission from the dynamic target was collected using a $4\times$ objective lens (NA 0.1), spectrally filtered by a dichroic mirror and an emission filter, and recorded by an sCMOS camera over $M\approx200$ frames.

Figure~\ref{fig4_incoherent}B shows a representative raw frame captured through the scattering layer, exhibiting severe scattering distortions.
Similarly to the coherent imaging experiments, before scattering compensation processing, we subtracted consecutive frames to suppress correlated, slowly varying fluctuations that violate the uncorrelated fluctuations condition. 
Applying the I-CLASS reconstruction algorithm to the preprocessed dataset yielded the corrected object frame shown in Fig.~\ref{fig4_incoherent}C and the reconstructed incoherent PSF shown in Fig.~\ref{fig4_incoherent}D. 
Consecutive corrected frames clearly reveal the motion of the microscopic beads along the tube (Fig.~\ref{fig4_incoherent}E), demonstrating time-resolved fluorescence imaging through the scattering layer. The complete reconstructed video sequence is provided in Supplementary Movies S2 and S3.

\section*{Discussion}

We have demonstrated that target dynamics, including \textit{in-vivo} blood flow, can serve as an intrinsic information source for computational scattering correction in coherent and incoherent imaging modalities. While target dynamics, such as blood flow or fluorescence blinking, have been previously exploited as a source of diversity for super-resolution imaging \cite{Chaigne2017, Dertinger2009}, here we apply this principle to scattering correction.
Beyond the fundamental general principle, a practical advantage of this approach is that it enables reconstruction of dynamic objects without requiring object stationarity during multiple acquisitions. This brings the acquisition time requirement to a single frame per dynamics correlation time, orders of magnitude reduction from the requirements of matrix-based methods \cite{weinberg2024fluorescence}. This approach also eliminates the need for illumination control or a spatial light modulator (SLM) and can be applied to widefield microscopes with no change in hardware.

The effectiveness of our method critically depends on the assumption that object motion generates sufficient spatial randomness over time. Two scenarios can compromise this requirement. First, large objects may evolve too slowly between consecutive frames, causing temporal pixel correlations. A potential solution is to increase the temporal interval between selected frames. Second, quasi-static components such as support structures or background can introduce persistent temporal correlations. We addressed this second challenge through consecutive-frame subtraction (Fig.~\ref{fig2_tube} and Fig.~\ref{fig4_incoherent}), which acts as a high-pass temporal filter isolating rapid sample-related fluctuations from slower system-related variations. More advanced spatiotemporal filtering approaches may also be applied \cite{Errico2015}.
An alternative strategy is to supplement the measurements with rapidly varying speckle illuminations.

For objects extending beyond a single isoplanatic region, we successfully employed a mosaicking approach by subdividing the FoV into isoplanatic patches (Figs.~\ref{fig2_tube} and \ref{fig3_finger}). This strategy is limited to cases where different patches remain separable from one another, a characteristic typically associated with thin scattering media. In thick scattering media, where this separability assumption breaks down, a multi-layer multi-conjugate correction may be applied \cite{kang2023tracing, Haim2025}.

A fundamental limitation of our framework is the assumption of a static scattering medium, an idealization that is often violated in practice, since most natural scattering media exhibit temporal dynamics. This constraint may be relaxed when the timescales of medium and object dynamics differ, allowing the slower-changing component to be treated as quasi-static. The critical challenge emerges when both evolve at comparable rates, rendering their contributions inseparable. This defines the operational boundary of our approach: it bridges two complementary regimes: static objects within dynamic media, where medium dynamics can provide the information diversity \cite{sunray2025matrix}, and dynamic objects within static media, where target dynamics serve this role.


\section*{Materials and Methods}

\subsubsection*{Experimental setup}
Fig.~\ref{fig2_tube} illustrates the experimental setup for coherent imaging of flowing absorbing targets through scattering media, based on off-axis holography to record complex field holograms. Illumination was provided by a 21~mW continuous-wave (CW) polarized He-Ne laser (HNL210L, Thorlabs) operating at 632.8~nm. A polarizing beam splitter (PBSW-633, Thorlabs) divided the beam into object and reference arms, with their optical path length difference maintained within the laser's coherence length (20~cm). Half-wave plates (WPH 10ME-633, Thorlabs) and polarizers (LPVISE-100-A, Thorlabs) in both arms enabled intensity control and polarization alignment.
In the object arm, a demagnifying telescope ($f=75$~mm, LA1608-A, and $f=25.4$~mm, LB1761-A; both from Thorlabs) reduced the beam diameter, focusing more optical power onto the target. A polarizing beam splitter (PBSW-633, Thorlabs) followed by a quarter-wave plate (WPQ 10ME-633 Ø1", Thorlabs) separated the illumination from the object back-reflection .

The back-reflected light from the target propagated through the diffuser toward the detector. A $4f$ imaging system with 1.6$\times$ magnification ($f=125$~mm, LA1384-A and $f=200$~mm, AC508-200-A-ML; both from Thorlabs) imaged the diffuser plane onto the CCD sensor (8051M-USB, Thorlabs). The object and reference beams were recombined at the camera using a non-polarizing beam splitter (BS03, Thorlabs). A 1~nm bandwidth band-pass filter (MaxLine Laser-line Filter  633, centered at 632.8~nm) was placed in front of the camera to isolate the illumination wavelength.

The object consisted of 6~$\mu$m optically absorbing polystyrene microspheres (Polybead, catalog no. 15714-5) flowing through a transparent silicone tube (Quickun Pure Silicone Tubing, 0.3~mm inner diameter, and 0.8~mm outer diameter). The tube was placed inside a water-filled cuvette, 5.3~cm behind a static 0.5$^\circ$ holographic diffuser (Newport).

In Fig.~\ref{fig3_finger}, the experimental setup was similar but with a different target. A negative USAF resolution target (R1DS1N, Groups 2–7, Ø1" Thorlabs) was placed approximately 7~cm behind the diffuser, with a human finger positioned farther behind as a dynamic object. The demagnifying telescope in the object arm was omitted in this configuration.

Fig.~\ref{fig4_incoherent} shows the fluorescence experimental configuration. The setup consists of a 488~nm continuous-wave (CW) laser (06-MLD-488, Cobolt) operated at 20~mW and a rapidly rotating holographic diffuser (EPS 0.5$^\circ$) positioned approximately 15~cm from the objective lens to produce quasi-uniform illumination. Image acquisition was performed using an Andor Neo 5.5 sCMOS camera integrated into a $4f$ imaging system equipped with a Thorlabs LA1256-A tube lens (focal length: 180~mm) and an Olympus 4$\times$ PLN objective (NA 0.1, WD 18.5~mm), providing a depth of field (DoF) of 50~$\mu$m. The fluorescence signal was spectrally filtered from the excitation illumination by a dichroic mirror (DMLP505R, Thorlabs) and an emission filter (MF525-39, Thorlabs).
The dynamic sample was positioned at the focal plane of the objective lens, with a holographic diffuser of $0.5^\circ$ (Newport) placed $\sim$5~mm in front of the objective and 13.5~mm from the sample.

The dynamic target consisted of fluorescent microspheres (Fluoresbrite YG, 10~$\mu$m) suspended in a solution with high concentration variability (c.v. $\gg 10\%$). The beads flowed through a transparent silicone tube with an inner diameter of 0.3~mm and an outer diameter of 0.8~mm. The tube was placed inside a 5~mm path-length glass cuvette filled with water. The flow was driven by a random manual pressing of the syringes connected to the tube.

\subsubsection*{Experimental Parameters}

The experimental parameters for the results presented in Figs.~\ref{fig2_tube}, \ref{fig3_finger}, and \ref{fig4_incoherent} are summarized below, including camera exposure times, image resolutions, and algorithm computation times.

For Fig.~\ref{fig2_tube}, holograms were captured with 5~ms exposure time at a frame rate of 17~fps and a resolution of 2472~$\times$~2472 pixels. Complex field retrieval was performed using a 941~$\times$~941 pixel crop in the Fourier domain (Fig.~\ref{fig2_tube}A), followed by further cropping to a 630~$\times$~630 pixel object region shown in Fig.~\ref{fig2_tube}B,C. The reconstructions shown in Fig.~\ref{fig2_tube}D,E,G,H used 9 patches of 230~$\times$~230 pixels with 30-pixel overlap between adjacent patches. Each patch was zero-padded to 690~$\times$~690 pixels before reconstruction, yielding correction phase masks of the same size ( Fig.~\ref{fig2_tube}F,I).

For Fig.~\ref{fig3_finger}, holograms were recorded with a 3~ms exposure time at a frame rate of 17~fps and a resolution of 2472~$\times$~2472 pixels. Complex field retrieval was performed using a 943~$\times$~943 pixel crop in the Fourier domain (Fig.~\ref{fig3_finger}A), followed by further cropping to a 660~$\times$~660 pixel object region for the demonstrations shown in Fig.~\ref{fig3_finger}B--D. The reconstruction employed manual subdivision into 12 patches of varying sizes corresponding to different object regions, with all pixels outside each patch set to zero within the 943~$\times$~943 pixel field.  Figure~\ref{fig3_finger}E shows the corresponding reconstructed phase masks over circular regions of approximately 300-pixel diameter.

For Fig.~\ref{fig4_incoherent}, 300 raw frames were captured at 1000~$\times$~1000 pixels with 15~ms exposure time at a frame rate of 30.9~fps. The compensation algorithm was applied to cropped images of size 450~$\times$~600 pixels. For display, images were further cropped to 450~$\times$~450 pixels for Fig.~\ref{fig4_incoherent}A, 300~$\times$~300 pixels for Fig.~\ref{fig4_incoherent}B,C, 350~$\times$~350 pixels for Fig.~\ref{fig4_incoherent}D, and 150~$\times$~150 pixels for Fig.~\ref{fig4_incoherent}E. Insets show 40~$\times$~40 pixel regions in Fig.~\ref{fig4_incoherent}B,C and 16~$\times$~16 pixel regions in Fig.~\ref{fig4_incoherent}E. 

The experiments shown in Figs.~\ref{fig2_tube} and \ref{fig3_finger} were reconstructed using $M=180$ realizations. For the incoherent imaging experiment shown in Fig.~\ref{fig4_incoherent}, frames in which isolated beads became visible in the reconstructed images after an initial reconstruction were excluded from the final dataset to rule out trivial reconstruction contributions. The final reconstruction was therefore performed using $M=198$ realizations (Supplementary Movie S2). Supplementary Movie S3 presents the reconstruction of the complete recorded sequence using the correction obtained from the reduced dataset. 

Runtime on GPU: I-CLASS (NVIDIA RTX 4090, 24~GB) required $\sim13$~s per patch for 200 iterations for Fig.~\ref{fig2_tube} (180 frames, 690~$\times$~690 pixels), $\sim57$~s per patch for 200 iterations for Fig.~\ref{fig3_finger} (180 frames, 943~$\times$~943 pixels), and $\sim6$~s for 200 iterations on the full field for Fig.~\ref{fig4_incoherent} (198 frames, 450~$\times$~600 pixels). Model-based computational wavefront shaping (NVIDIA GeForce RTX 3090) required $\sim735$~s per patch for 2000 iterations for Fig.~\ref{fig2_tube} (180 frames, 690~$\times$~690 pixels).

Note that, when displaying the fields in Fig.~\ref{fig2_tube}A,B,D,G and Fig.~\ref{fig3_finger}A,B using the HSV colormap (where the hue represents the phase and the value represents the amplitude), the amplitude channel was normalized to the 99th percentile of the amplitude distribution.

\subsubsection*{Coherent matrix-based reconstruction pre-processing}

To enhance the convergence of the matrix-based (CTR-CLASS) algorithm and relax the sampling constraints, we implemented a numerical defocus pre-compensation step. 
In the raw measurements, the free-space propagation from the target to the diffuser introduces a rapidly oscillating quadratic phase term (chirp) into the captured fields. Resolving these high-frequency oscillations typically necessitates a high sampling density. 
To mitigate this, we first estimated the approximate axial distance, $z_{\text{est}}$, between the object and the scattering layer. This was achieved by numerically back-propagating the captured aberrated fields and identifying the distance yielding maximal approximate focus, despite the severe scattering distortions. 
Subsequently, we multiplied the recorded fields $E_\text{m}(\vec{r})$ by a conjugate quadratic phase factor corresponding to this estimated distance, obtaining the quadratic-phase-compensated fields: 
\begin{equation}
    E'_\text{m}(\vec{r}) = E_\text{m}(\vec{r}) \cdot \exp\left(-i \frac{k}{2 z_{\text{est}}} |\vec{r}|^2\right)
\end{equation}
\noindent 
where $k = 2 \pi /\lambda$ is the illumination wavenumber. 
This preconditioning effectively removes the bulk of the diffractive curvature from the data. Consequently, the phase mask retrieved by the CTR-CLASS algorithm represents only the residual phase contributions: the aberrations of the scattering medium and the fine-tuning of the propagation distance ($\Delta z = z_{\text{true}} - z_{\text{est}}$). This residual phase exhibits significantly lower spatial frequencies than the full propagation kernel, thereby reducing the computational complexity and improving the stability of the reconstruction.

\subsubsection*{Deconvolution in the incoherent imaging modality}

In the incoherent imaging modality, correcting only the phase of the optical transfer function (OTF) is insufficient, as the reconstructed image still contains residual haze and background contributions~\cite{weinberg2024fluorescence}. Therefore, after phase correction, the object was further corrected by compensating the OTF amplitude using a Wiener deconvolution~\cite{weinberg2024fluorescence}:

\begin{equation}\label{deconv}
    \tilde{A}_{\text{fixed}}(\vec{k}) 
    = \frac{\tilde{A}_\text{m}(\vec{k})}{\text{MTF}(\vec{k}
    ) + \sigma},
\end{equation}

\noindent
where $\tilde{A}_\text{m}(\vec{k})$ is the phase-corrected frame in Fourier space, $\text{MTF}(\vec{k})$ is the estimated modulation transfer function (see Supplementary Text  S2), and $\sigma$ is a regularization parameter.

For the incoherent result shown in Fig.~\ref{fig4_incoherent}, an additional low-pass filter was applied after Wiener deconvolution to remove spatial frequencies beyond the passband of the optical system.


\begin{figure}[H]
	\centering
	\includegraphics[width=0.9\textwidth]{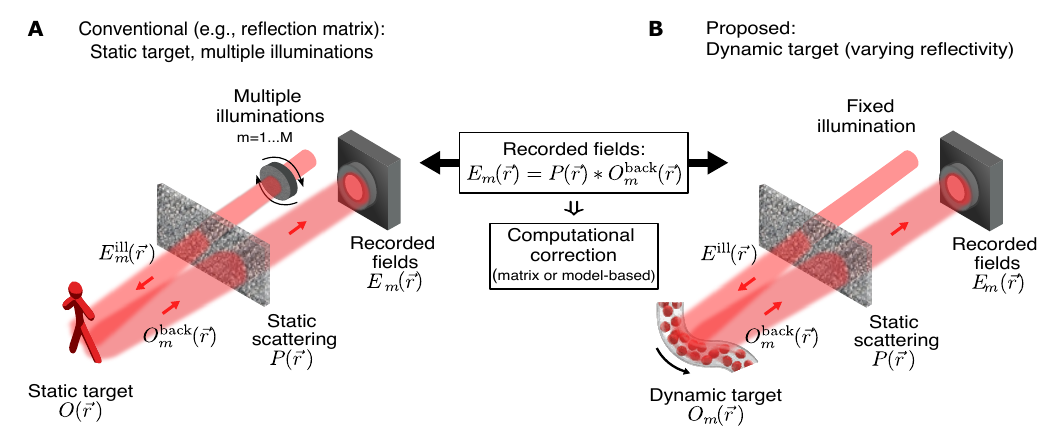}
\caption{\footnotesize
\textbf{Concept: The equivalence between matrix-based imaging of static targets through scattering media using multiple illuminations,  and the proposed dynamic-targets imaging.}
(\textbf{A}) Conventional (e.g. reflection-matrix based) imaging of static targets through scattering media using multiple illuminations: the object reflectance $O(\vec{r})$ remains constant, while the illumination patterns vary temporally to enable the acquisition of $M$ fields, $E^{\text{ill}}_\text{m}(\vec{r})$, $m=1..M$, that constitute the sample reflection-matrix. The fields reflected from the target plane are $O^{\text{back}}_\text{m}(\vec{r}) = O(\vec{r})\,E^{\text{ill}}_\text{m}(\vec{r})$, where $E^{\text{ill}}_\text{m}(\vec{r})$ are the varying illuminating fields at the target plane. 
(\textbf{B}) The proposed approach for imaging dynamic targets through scattering media: multiple acquisitions of the field reflected from the target, while the target is dynamically varying are captured. The target reflectance, $O_\text{m}(\vec{r})$, varies temporally ($m=1..M$) between acquisitions, while the illumination, $E^{\text{ill}}(\vec{r})$, remains fixed. The back-reflected field at the target plane is: $O^{\text{back}}_\text{m}(\vec{r}) = O_\text{m}(\vec{r})\,E^{\text{ill}}(\vec{r})$. 
In both scenarios (A,B), the captured fields $E_\text{m}(\vec{r})$ have the same algebraic structure, which is (in the isoplanatic case): $E_\text{m}(\vec{r}) = P(\vec{r}) \ast O^{\text{back}}_\text{m}(\vec{r})$, where $P(\vec{r})$ is the medium's scattering PSF.
The equivalence also holds for the case of incoherent and anisoplanatic imaging (see text). This equivalence enables applying the same computational correction approach to datasets from either case.}
  \label{fig1_principle}
\end{figure}

\begin{figure}[H]
\centering
	 \includegraphics[width=0.9\textwidth]{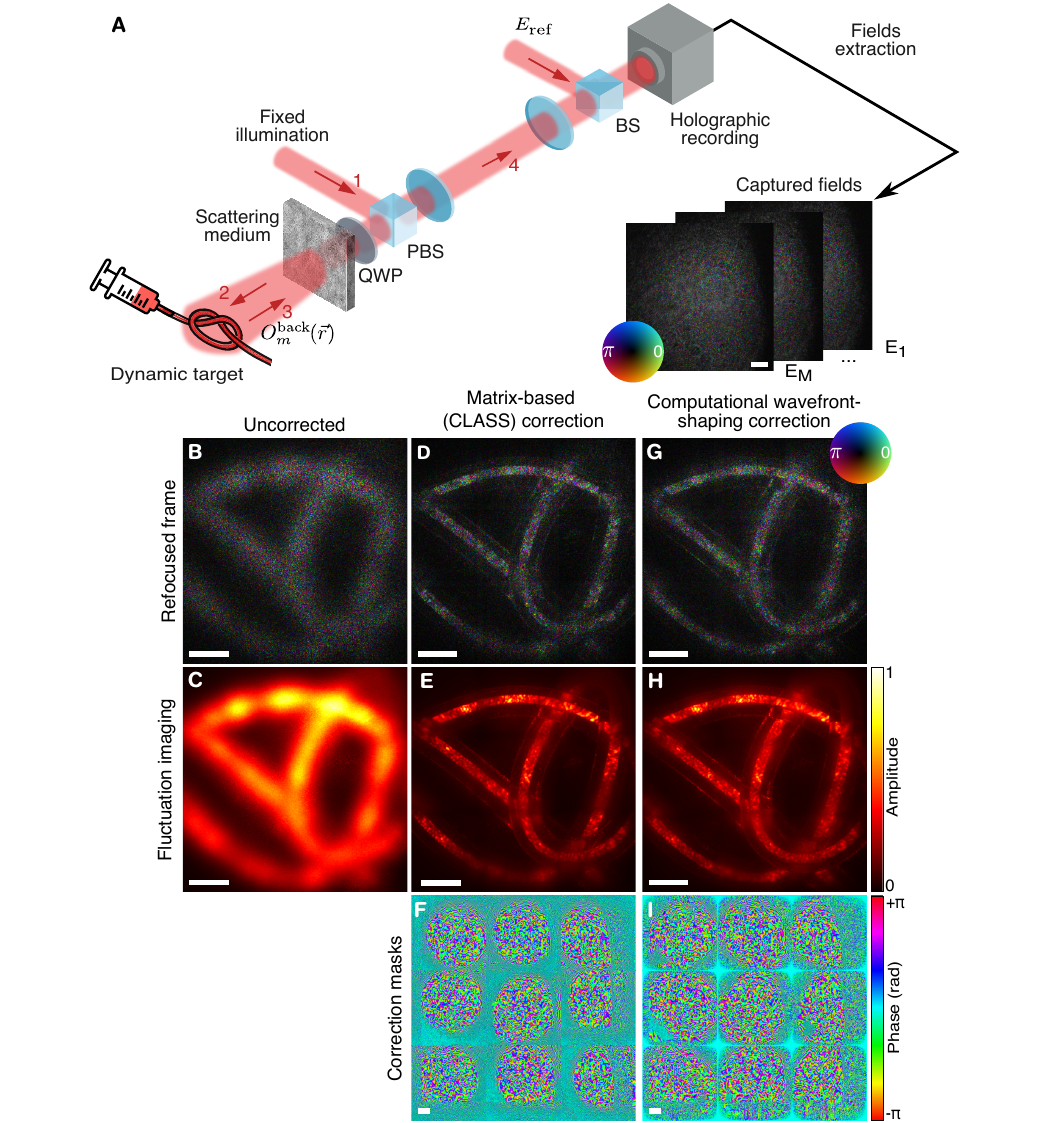}

\caption{\footnotesize
\textbf{Coherent imaging of flowing beads through a scattering layer.} 
(\textbf{A}) Experimental setup: A dynamic target of polystyrene microspheres flowing in a transparent tube was placed behind a scattering diffuser. Fixed coherent illumination is provided through the diffuser. A series of $180$ recorded fields, $E_\text{m}(\vec{r})$, $m=1..180$, are holographically recorded in an epi-detection geometry by a CCD camera imaging the scattering medium surface, via off-axis holography.
(\textbf{B}) An image of one captured field back-propagated to the target plane without scattering correction. 
(\textbf{C}) Fluctuation image (temporal standard deviation) of the fields at the target plane without scattering correction. (\textbf{D,E}) Same as (B,C) but after applying matrix-based correction (CTR-CLASS \cite{lee22, weinberg2024fluorescence}): 
(\textbf{F}) The recovered scattering compensation phase masks at nine sub-regions of the field of view account for anisoplanatic scattering. (\textbf{G-I}) Image-guided computational wavefront-shaping correction applied on the same dataset\cite{Haim2025}, presented in the same manner as D-F. Both scattering compensation schemes successfully undo scattering and reconstruct a video of the target dynamics. Scale bar: 1~mm. PBS, polarizing beam splitter; BS, beam splitter; QWP, quarter-wave plate.}
  \label{fig2_tube}
\end{figure}

\begin{figure}[H]
	\centering
	\includegraphics[width=0.9\textwidth]{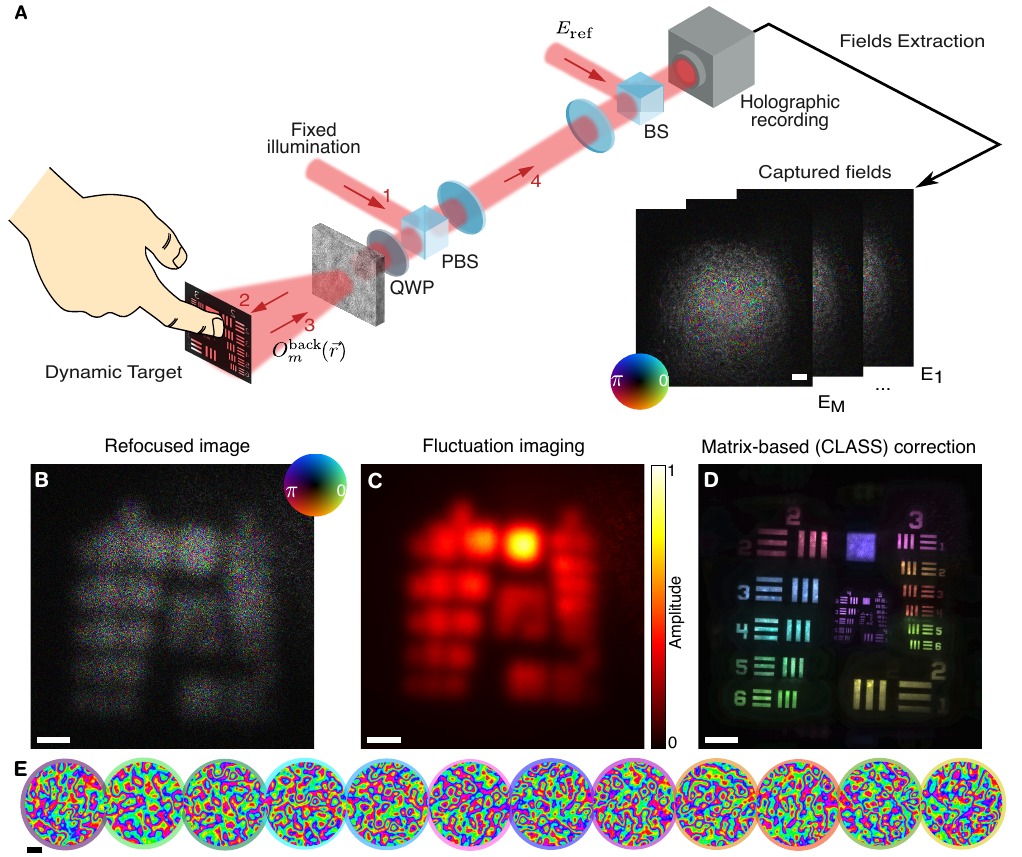}
\caption{\footnotesize
\textbf{Exploiting \textit{in-vivo} blood-flow dynamics for scattering compensation.}
(\textbf{A}) Experimental setup: a human finger placed behind a negative USAF target provides naturally rapidly varying reflectivity dynamics. The target is illuminated by a fixed laser illumination through a scattering diffuser. A CCD camera captured $M=180$ back-scattered complex holographic fields in reflection geometry.
(\textbf{B}) An image of one captured field back-propagated to the target plane without scattering correction. 
(\textbf{C}) Fluctuation image (temporal standard deviation) of the fields at the target plane without scattering correction.
(\textbf{D}) Same as (C) after applying CTR-CLASS \cite{lee22,weinberg2024fluorescence} on 12 different patches of the field of view (color-coded in D). (\textbf{E}) The recovered scattering compensation phase masks at the 12 sub-regions of the field of view.
Scale bar: 1~mm. PBS, polarizing beam splitter; BS, beam splitter; QWP, quarter-wave plate.}
\label{fig3_finger}
\end{figure}

\begin{figure}[H]
	\centering
	\includegraphics [width=0.9\textwidth]{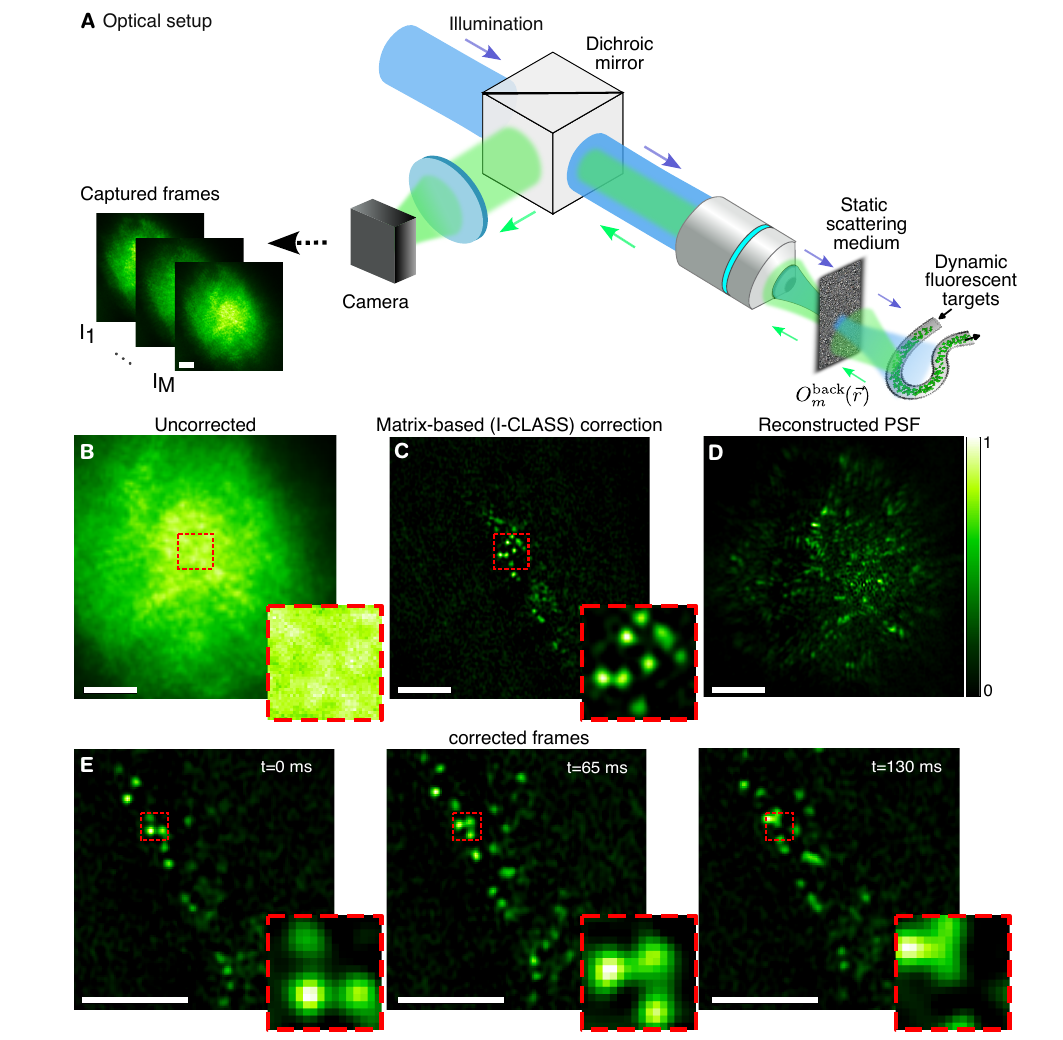}
\caption{\footnotesize
\textbf{Widefield fluorescence microscopy of flowing beads through a scattering layer.} 
(\textbf{A}) Experimental setup: a dynamic target of 10~$\mu$m diameter fluorescent beads flowing in a transparent tube was placed inside a water-filled cuvette behind a static scattering diffuser. Homogeneous illumination is provided through the diffuser. An sCMOS camera captures a video of $M\approx200$ frames of the scattered fluorescence in epi-detection geometry.
(\textbf{B}) A single raw camera frame, displaying the strong scattering distortion leading to a low-contrast, featureless image (a zoomed-in region is shown in the inset).
(\textbf{C}) Same as (B) after applying matrix-based I-CLASS algorithm\cite{weinberg2024fluorescence} on the captured dataset. The corrected frame reveals the individual flowing beads. \textbf{(D)} The I-CLASS reconstructed scattering PSF. 
(\textbf{E}) Several consecutive frames from the scattering-corrected video revealing the beads' motion (see Supplementary Movie S2 and S3 for the full sequence).
Scale bar: 0.1~mm.}
\label{fig4_incoherent}
\end{figure}


\clearpage 
\bibliography{bib}

@Article{Monin2025,
author={Monin, Sagi
and Alterman, Marina
and Levin, Anat},
title={Rapid wavefront shaping using an optical gradient acquisition},
journal={Nature Communications},
year={2026},
month={Jan},
day={10},
volume={17},
number={1},
pages={1537},
issn={2041-1723},
doi={10.1038/s41467-025-68259-2},
}

@Article{Aizik2024,
title={Non-invasive and noise-robust light focusing using confocal wavefront shaping},
author={Aizik, Dror
and Levin, Anat},
journal={Nature Communications},
year={2024},
month={Jul},
day={02},
volume={15},
number={1},
pages={5575},
issn={2041-1723},
doi={10.1038/s41467-024-49697-w},
}

@article{badon2020distortion,
  title={Distortion matrix concept for deep optical imaging in scattering media},
  author={Badon, Amaury and Barolle, Victor and Irsch, Kristina and Boccara, A Claude and Fink, Mathias and Aubry, Alexandre},
  journal={Science Advances},
  volume={6},
  number={30},
  pages={eaay7170},
  year={2020},
  doi = {10.1126/sciadv.aay7170},
  publisher={American Association for the Advancement of Science}
}

@Article{Kang2017,
author={Kang, Sungsam
and Kang, Pilsung
and Jeong, Seungwon
and Kwon, Yongwoo
and Yang, Taeseok D.
and Hong, Jin Hee
and Kim, Moonseok
and Song, Kyung--Deok
and Park, Jin Hyoung
and Lee, Jun Ho
and Kim, Myoung Joon
and Kim, Ki Hean
and Choi, Wonshik},
title={High-resolution adaptive optical imaging within thick scattering media using closed-loop accumulation of single scattering},
journal={Nature Communications},
year={2017},
month={Dec},
day={18},
volume={8},
number={1},
pages={2157},
issn={2041-1723},
doi={10.1038/s41467-017-02117-8},

}

@article{lee22,
  title={High-throughput volumetric adaptive optical imaging using compressed time-reversal matrix},
  author={Lee, Hojun and Yoon, Seokchan and Loohuis, Pascal and Hong, Jin Hee and Kang, Sungsam and Choi, Wonshik},
  journal={Light: Science \& Applications},
  volume={11},
  number={1},
  pages={16},
  year={2022},
  publisher={Nature Publishing Group UK London},
  doi = {10.1038/s41377-021-00705-4}
}

@article{feng23,
  title={NeuWS: Neural wavefront shaping for guidestar-free imaging through static and dynamic scattering media},
  author={Brandon Y. Feng  and Haiyun Guo  and Mingyang Xie  and Vivek Boominathan  and Manoj K. Sharma  and Ashok Veeraraghavan  and Christopher A. Metzler },
  journal={Science Advances},
  volume={9},
  number={26},
  pages={eadg4671},
  year={2023},
  doi = {10.1126/sciadv.adg4671},
  publisher={American Association for the Advancement of Science}
}

@book{goodman2005introduction,
  title={Introduction to Fourier optics},
  author={Goodman, Joseph W},
  year={2005},
  publisher={Roberts and Company publishers}
}

@article{bertolotti2022imaging,
  title={Imaging in complex media},
  author={Bertolotti, Jacopo and Katz, Ori},
  journal={Nature Physics},
  volume={18},
  number={9},
  pages={1008--1017},
  year={2022},
  publisher={Nature Publishing Group UK London},
  issn={1745-2481},
  doi={10.1038/s41567-022-01723-8}
}

@article{kang2023tracing,
  title={Tracing multiple scattering trajectories for deep optical imaging in scattering media},
  author={Kang, Sungsam and Kwon, Yongwoo and Lee, Hojun and Kim, Seho and Hong, Jin Hee and Yoon, Seokchan and Choi, Wonshik},
  journal={Nature Communications},
  volume={14},
  year={2023},
  doi = {10.1038/s41467-023-42525-7}
}

@article{yeminy2021guidestar,
  title={Guidestar-free image-guided wavefront shaping},
  author={Yeminy, Tomer and Katz, Ori},
  journal={Science advances},
  volume={7},
  number={21},
  pages={eabf5364},
  doi = {10.1126/sciadv.abf5364},
  year={2021},
  publisher={American Association for the Advancement of Science}
}

@article{freund1988memory,
  title={Memory effects in propagation of optical waves through disordered media},
  author={Freund, Isaac and Rosenbluh, Michael and Feng, Shechao},
  journal={Physical review letters},
  volume={61},
  number={20},
  pages={2328},
  year={1988},
  publisher={APS},
  DOI ={https://doi.org/10.1103/PhysRevLett.61.238}
}

@article{Chaigne2017,
  author    = {Chaigne, Thomas and Arnal, Bastien and Vilov, Sergey and Bossy, Emmanuel and Katz, Ori},
  title     = {Super-resolution photoacoustic imaging via flow-induced absorption fluctuations},
  journal   = {Optica},
  volume    = {4},
  number    = {11},
  pages     = {1397--1404},
  year      = {2017},
  publisher = {Optica Publishing Group},
  doi       = {10.1364/OPTICA.4.001397}
}

@article{wu2016single,
  title={Single-shot diffraction-limited imaging through scattering layers via bispectrum analysis},
  author={Wu, Tengfei and Katz, Ori and Shao, Xiaopeng and Gigan, Sylvain},
  journal={Optics Letters},
  volume={41},
  number={21},
  pages={5003--5006},
  year={2016},
  publisher={Optica Publishing Group},
  doi = {10.1364/OL.41.005003}
}

@article{ventalon2005quasi,
  title={Quasi-confocal fluorescence sectioning with dynamic speckle illumination},
  author={Ventalon, Cathie and Mertz, Jerome},
  journal={Optics letters},
  volume={30},
  number={24},
  pages={3350--3352},
  year={2005},
  publisher={Optica Publishing Group},
  doi = {10.1364/OL.30.003350}
}

@article{fienup1978reconstruction,
  title={Reconstruction of an object from the modulus of its Fourier transform},
  author={Fienup, James R},
  journal={Optics letters},
  volume={3},
  number={1},
  pages={27--29},
  year={1978},
  publisher={Optica Publishing Group},
  doi = {10.1364/OL.3.000027}
}

@inproceedings{xie2024wavemo,
  title={WaveMo: Learning Wavefront Modulations to See Through Scattering},
  author={Xie, Mingyang and Guo, Haiyun and Feng, Brandon Y and Jin, Lingbo and Veeraraghavan, Ashok and Metzler, Christopher A},
  booktitle={Proceedings of the IEEE/CVF Conference on Computer Vision and Pattern Recognition},
  pages={25276--25285},
  year={2024},
  doi = {10.1109/CVPR52733.2024.02388}
}

@article{weinberg2024fluorescence,
  title={Noninvasive megapixel fluorescence microscopy through scattering layers by a virtual incoherent reflection matrix},
  author={Weinberg, Gil and Sunray, Elad and Katz, Ori},
  journal={Science Advances},
  volume={10},
  number={47},
  pages={eadl5218},
  year={2024},
  publisher={American Association for the Advancement of Science},
  doi = {10.1126/sciadv.adl5218}
}

@Article{Haim2025,
author={Haim, Omri and Boger-Lombard, Jeremy and Katz, Ori},
title={Image-guided computational holographic wavefront shaping},
journal={Nature Photonics},
year={2025},
month={Jan},
day={01},
volume={19},
number={1},
pages={44-53},
issn={1749-4893},
doi={10.1038/s41566-024-01544-6},

}

@article{gigan2022roadmap,
  title={Roadmap on wavefront shaping and deep imaging in complex media},
  author={Gigan, S. and others},
  journal={Journal of Physics: Photonics},
  DOI = {10.1088/2515-7647/ac76f9},
  volume={4},
  pages={042501},
  year={2022}
}

@article{lee2023exploiting,
  title={Exploiting volumetric wave correlation for enhanced depth imaging in scattering medium},
  author={Lee, Y.-R. and Kim, D.-Y. and Jo, Y. and Kim, M. and Choi, W.},
  journal={Nature Communications},
  volume={14},
  year={2023},
  doi = {10.1038/s41467-023-37467-z}
}

@article{Balondrade2024,
  author = {Balondrade, Paul and Barolle, Victor and Guigui, Nicolas and Auriant, Emeric and Rougier, Nathan and Boccara, Claude and Fink, Mathias and Aubry, Alexandre},
  title = {Multi-spectral reflection matrix for ultrafast 3D label-free microscopy},
  journal = {Nature Photonics},
  year = {2024},
  volume = {18},
  number = {10},
  pages = {1097--1104},
  doi = {10.1038/s41566-024-01479-y},
}

@article{katz2014non,
  title={Non-invasive single-shot imaging through scattering layers and around corners via speckle correlations},
  author={Katz, Ori and Heidmann, Pierre and Fink, Mathias and Gigan, Sylvain},
  journal={Nature Photonics},
  volume={8},
  number={10},
  pages={784--790},
  year={2014},
  doi = {10.1038/nphoton.2014.189},
 publisher={Nature Publishing Group UK London}
}

@article{labeyrie1970attainment,
  title={Attainment of diffraction limited resolution in large telescopes by Fourier analysing speckle patterns in star images},
  author={Labeyrie, A.},
  journal={Astronomy and Astrophysics},
  volume={6},
  pages={85--87},
  year={1970},
}

@article{ayers1988knox,
  title={Knox-Thompson and triple-correlation imaging through atmospheric turbulence},
  author={Ayers, G. and Northcott, M. and Dainty, J.},
  journal={Journal of the Optical Society of America A},
  volume={5},
  number={7},
  pages={963--985},
  month = {Jul},
  year={1988},
  publisher = {Optica Publishing Group},
  doi = {10.1364/JOSAA.5.000963}
}

@article{vellekoop2007focusing,
  title={Focusing coherent light through opaque strongly scattering media},
  author={Vellekoop, I.M. and Mosk, A.P.},
  journal={Optics Letters},
  volume={32},
  pages={2309--2311},
  year={2007}, 
  doi = {10.1364/OL.32.002309},
}

@article{sunray2025matrix,
  author    = {Elad Sunray and Gil Weinberg and Benzy Laufer and Ori Katz},
  title     = {Matrix-based imaging through dynamic scattering},
  journal   = {Nature Communications},
  year      = {2025},
  volume    = {16},
  number    = {1},
  pages     = {9413},
  doi       = {10.1038/s41467-025-64422-x}
}

@Article{gupta2025imaging,
      title={Imaging through volumetric scattering media by decoding angular light paths}, 
      author={Kalpak Gupta and Dinh Hoang Tran and Sungsam Kang and Yongwoo Kwon and Seokchan Yoon and Jin Hee Hong and Ye-Ryoung Lee and Wonshik Choi},
      year={2025},
      journal={arXiv e-prints},
      doi = {10.48550/arXiv.2509.11491}
}

@article{Zhang2025Adaptive,
author = {Yiwen Zhang and Minh Dinh and Zeyu Wang and Tianhao Zhang and Tianhang Chen and Chia Wei Hsu},
title = {{Adaptive optical multispectral matrix approach for label-free high-resolution imaging through complex scattering media}},
volume = {7},
journal = {Advanced Photonics},
number = {4},
publisher = {SPIE},
pages = {046008},
year = {2025},
doi = {10.1117/1.AP.7.4.046008},

}

@article{Snieder2010,
  author = {Snieder, Roel and Wapenaar, Kees},
  title = {Imaging with ambient noise},
  journal = {Physics Today},
  volume = {63},
  number = {9},
  pages = {44--49},
  year = {2010},
  month = {September},
  doi = {10.1063/1.3490500}
}

@article{Farah2024,
author = {Yusef Farah and Gabe Murray and Jeff Field and Maxine Varughese and Lang Wang and Olivier Pinaud and Randy Bartels},
journal = {Optica},
number = {5},
pages = {693--705},
publisher = {Optica Publishing Group},
title = {Synthetic spatial aperture holographic third harmonic generation microscopy},
volume = {11},
month = {May},
year = {2024},

doi = {10.1364/OPTICA.521088},
}

@article{Kang2025,
  author  = {Kang, Sungsam and Yoon, Seokchan and Choi, Wonshik},
  title   = {Implementation of reflection matrix microscopy: an algorithm perspective},
  journal = {Journal of Physics: Photonics},
  year    = {2025},
  volume  = {7},
  number  = {1},
  doi     = {10.1088/2515-7647/adb7af},
}

@article{Jiang2026,
author = {Jiang, Weiyun and Guo, Haiyun and Metzler, Christopher A. and Veeraraghavan, Ashok},
title = {Guidestar-Free Adaptive Optics with Asymmetric Apertures},
year = {2026},
publisher = {Association for Computing Machinery},
address = {New York, NY, USA},
issn = {0730-0301},
doi = {10.1145/3809501},
note = {Just Accepted},
journal = {ACM Trans. Graph.},
month = apr
}

@article{Qureshi2017,
author = {Muhammad Mohsin Qureshi and Joshua Brake and Hee-Jae Jeon and Haowen Ruan and Yan Liu and Abdul Mohaimen Safi and Tae Joong Eom and Changhuei Yang and Euiheon Chung},
journal = {Biomed. Opt. Express},
number = {11},
pages = {4855--4864},
publisher = {Optica Publishing Group},
title = {In vivo study of optical speckle decorrelation time across depths in the mouse brain},
volume = {8},
month = {Nov},
year = {2017},
doi = {10.1364/BOE.8.004855},
}

@article{Dertinger2009
,
author = {T. Dertinger  and R. Colyer  and G. Iyer  and S. Weiss  and J. Enderlein },
title = {Fast, background-free, 3D super-resolution optical fluctuation imaging (SOFI)},
journal = {Proceedings of the National Academy of Sciences},
volume = {106},
number = {52},
pages = {22287-22292},
year = {2009},
doi = {10.1073/pnas.0907866106}}

@Article{Errico2015,
author={Errico, Claudia
and Pierre, Juliette
and Pezet, Sophie
and Desailly, Yann
and Lenkei, Zsolt
and Couture, Olivier
and Tanter, Mickael},
title={Ultrafast ultrasound localization microscopy for deep super-resolution vascular imaging},
journal={Nature},
year={2015},
month={Nov},
day={01},
volume={527},
number={7579},
pages={499-502},
issn={1476-4687},
doi={10.1038/nature16066},
}
\bibliographystyle{sciencemag}


\section*{Acknowledgments}

The authors thank Yevgeny Slobodkin and Jeremy Boger-Lombard for their helpful comments on the writing and presentation of this manuscript.

 \paragraph*{Funding:}
This project was supported by the H2020 European Research Council (101002406).

\paragraph*{Author contributions:}
O.K. proposed and conceptualized the project; I.T, B.L., and O.K. designed the coherent imaging experimental setup. Y.B. and O.K. designed the incoherent imaging experimental setup. I.T. carried out the coherent imaging experiments.
I.T, O.H., and Y.B. performed the coherent imaging data analysis;
Y.B. carried out the incoherent imaging measurements and data analysis;  
O.K. supervised the project.
O.K., Y.B, and I.T wrote the manuscript. 
All authors reviewed and approved the manuscript. 

\paragraph*{Ethics:}
This study involved the acquisition of optical measurements from a human volunteer's finger, which served as the dynamic temporal fluctuations required for the experiment presented in Fig.~\ref{fig3_finger}. The study was reviewed and approved by the Institutional Review Board (IRB) for Non-Medical Human Subjects Research of the Hebrew University of Jerusalem  (IRB Protocol No. 18022026, approved 18 February 2026). Informed consent was obtained from the participant after the nature and possible consequences of the study were explained. 

\paragraph*{Competing interests:}
The authors declare no competing interests.

\paragraph*{Data and materials availability:}
All data needed to evaluate the conclusions in the paper are present in the paper and/or the Supplementary Materials. The experimental data displayed in the figures are openly accessible at https://doi.org/10.5281/zenodo.20578800.

\subsection*{Supplementary materials}
Supplementary Text S1 to S5\\
Figure S1\\
Captions for Movies S1 to S3\\


\newpage


\renewcommand{\thefigure}{S\arabic{figure}}
\renewcommand{\thetable}{S\arabic{table}}
\renewcommand{\theequation}{S\arabic{equation}}
\renewcommand{\thepage}{S\arabic{page}}
\setcounter{figure}{0}
\setcounter{table}{0}
\setcounter{equation}{0}
\setcounter{page}{1} 

\begin{center}
\section*{Supplementary Materials for\\ \scititle}

Yoav~Ben~Haim$^{\dagger}$,
Ilay~Tomer$^{\dagger}$,
Omri~Haim,
Benzy~Laufer,
Ori~Katz$^{\ast}$\\
\small$^\ast$Corresponding author. Email: orik@mail.huji.ac.il\\
\small$^\dagger$These authors contributed equally to this work.
\end{center}

\subsubsection*{This PDF file includes:}
Supplementary Text S1 to S5 \\
Figure S1\\
Captions for Movies S1 to S3\\

\subsubsection*{Other Supplementary Materials for this manuscript:}
Movies S1 to S3\\

\newpage


\subsection*{S1: Equivalence of imaging a dynamic target to a static target with varying illumination in incoherent imaging}

In the case of incoherent imaging (e.g., fluorescence microscopy), the measured intensity at the detector plane is given by:
\begin{equation}\label{measure intensity integral}
I_\text{m}\!\left(\vec{r}\right) = \!\int\! P_\text{inc}\!\left(\vec{r}, \vec{r}\,'\right) 
\,\left[O_\text{m}\!\left(\vec{r}\,'\right) 
\, I^\text{ill}\!\left(\vec{r}\,'\right) \right]
\, d\vec{r}\,'
\end{equation}
\noindent
where $P_\text{inc}\!\left(\vec{r}, \vec{r}\,'\right)$ is the incoherent PSF describing intensity propagation through the scattering medium, $O_\text{m}\!\left(\vec{r}\,'\right)$ represents the temporally-varying fluorescence efficiency or reflectance of the target at time $t_\text{m}$, and $I^\text{ill}\!\left(\vec{r}\,'\right)$ is the temporally fixed illumination intensity at the target plane. Note that $P_\text{inc}$, the incoherent PSF, differs from $P$, the coherent PSF  used in Equation~\ref{measure field integral}, as it describes intensity (not field) propagation and is related to $P$ through $P_\text{inc} = |P|^2$ \cite{goodman2005introduction}.

Similar to the coherent case, this formulation for dynamic targets under fixed illumination is mathematically equivalent to static target imaging under time-varying illumination (with the temporal index $m$) :
\begin{equation}
I_\text{m}\!\left(\vec{r}\right) = \!\int\! P_\text{inc}\!\left(\vec{r}, \vec{r}\,'\right) 
\, \left[O\!\left(\vec{r}\,'\right) 
\, I_\text{m}^\text{ill}\!\left(\vec{r}\,'\right) \right]
\, d\vec{r}\,'
\end{equation}

This equivalence enables the application of matrix-based computational correction (I-CLASS~\cite{weinberg2024fluorescence}).

\subsection*{S2: Dynamic-target covariance matrix as a virtual reflection matrix}

We demonstrate the equivalence between the temporal covariance matrix of dynamic-target measurements and the virtual reflection matrix used in CTR-CLASS~\cite{lee22} and I-CLASS~\cite{weinberg2024fluorescence} for matrix-based scattering compensation.

\subsubsection*{Measurement matrix formulation}

In coherent holographic imaging of a dynamic target through a static scattering medium, the measured field at the detector plane at time $t_m$ is (Fig.~\ref{fig1_principle}B):
\begin{equation}
    E_m(\vec{r})
    = \int P(\vec{r}, \vec{r}')\,
      O_m^{\text{back}}(\vec{r}')\,
      d\vec{r}',
    \label{eq:Em}
\end{equation}
\noindent
where $P(\vec{r}, \vec{r}')$ is the coherent PSF of the static scattering medium, describing the propagation of light from point $\vec{r}'$ at the object plane to point $\vec{r}$ at the detector plane. For convenience, we define
\begin{equation}
    O_m^{\text{back}}(\vec{r}')
    = O_m(\vec{r}')\, E^{\text{ill}}(\vec{r}')
    \label{eq:Oback}
\end{equation}
\noindent
as the $m$-th realization of the back-reflected dynamic object field, where $O_m(\vec{r}')$ is the temporally-varying complex reflectance of the target at time $t_m$, and $E^{\text{ill}}(\vec{r}')$ is the temporally fixed illumination field at the target plane.

We acquire a sequence of $M$ such field measurements over time, each represented by $N$ spatial pixels. Reshaping each field into a column vector and stacking them yields the measurement matrix (Fig.~\ref{fig:S1_simulation}A):
\begin{equation}
    \mathbf{A} = \mathbf{P}\,\mathbf{O}^{\text{back}},
    \label{eq:A}
\end{equation}
\noindent
where $\mathbf{P} \in \mathbb{C}^{N \times N}$ is the static scattering matrix (the discretized form of $P(\vec{r},\vec{r}')$) and $\mathbf{O}^{\text{back}} \in \mathbb{C}^{N \times M}$ contains the $M$ temporal realizations of $O_m^{\text{back}}$ as its columns. Directly recovering the object from $\mathbf{A}$ is challenging: both $\mathbf{P}$ and $\mathbf{O}^{\text{back}}$ are unknown, and $\mathbf{P}$ transforms the object fields into seemingly uninformative speckle-like patterns (Fig.~\ref{fig:S1_simulation}A). 

\subsubsection*{Spatial decorrelation of object fluctuations}

We define the temporally fluctuating component of the object field as
\begin{equation}
    \hat{\mathbf{O}}
    \stackrel{\text{def}}{=}
    \mathbf{O}^{\text{back}}
    -
    \langle\mathbf{O}^{\text{back}}\rangle_{\text{row}},
    \label{eq:Ohat}
\end{equation}
\noindent
where $\langle\cdot\rangle_{\text{row}}$ denotes temporal averaging performed independently at each spatial location. Our framework relies on the assumption that temporal fluctuations arising from distinct spatial locations are approximately uncorrelated, as is the case, for example, in blood flow~\cite{Chaigne2017}:
\begin{equation}
    \hat{\mathbf{O}}_{i,:}\,\hat{\mathbf{O}}_{j,:}^{\dagger}
    \propto \delta_{i,j},
    \label{eq:decorr}
\end{equation}
\noindent\
where $\delta_{i,j}$ is the Kronecker delta, indicating that the temporal dynamics between different locations at the target plane are uncorrelated. 

\subsubsection*{Covariance matrix as a virtual reflection matrix}

Substituting Eqs.~(\ref{eq:A}) and~(\ref{eq:Ohat}) into the definition of the temporal covariance matrix, and invoking the spatial decorrelation assumption~(\ref{eq:decorr}), gives (Fig.~\ref{fig:S1_simulation}B):
\begin{equation}
    \mathbf{R}^{\text{virt}}
    \stackrel{\text{def}}{=}
    \text{Cov}(\mathbf{A})
    =
    \bigl(\mathbf{A} - \langle\mathbf{A}\rangle_{\text{row}}\bigr)
    \bigl(\mathbf{A} - \langle\mathbf{A}\rangle_{\text{row}}\bigr)^{\dagger}
    =
    \mathbf{P}\,\hat{\mathbf{O}}\hat{\mathbf{O}}^{\dagger}\,\mathbf{P}^{\dagger}
    \approx
    \mathbf{P}\,|\hat{\mathbf{O}}|^2\,\mathbf{P}^{\dagger},
    \label{eq:Rvirt}
\end{equation}
\noindent
where the approximation signifies that  $\hat{\mathbf{O}}\hat{\mathbf{O}}^{\dagger}$ becomes approximately diagonal under assumption~(\ref{eq:decorr}).  Equation~(\ref{eq:Rvirt}) becomes mathematically identical to the virtual reflection matrix arising in conventional matrix-based imaging of static objects under randomly varying illumination~\cite{lee22,weinberg2024fluorescence}, where the role of the varying illumination patterns is played by the temporally decorrelated object dynamics. In the isoplanatic regime, $\mathbf{P}$ and $\mathbf{P}^{\dagger}$ are convolution matrices, and $\mathbf{A}$ can therefore be processed using the same algorithmic pipeline as CTR-CLASS~\cite{lee22,weinberg2024fluorescence}.

\subsubsection*{Reconstruction pipeline and numerical validation}

\begin{figure}
\centering
\includegraphics [width=0.9\textwidth,]
{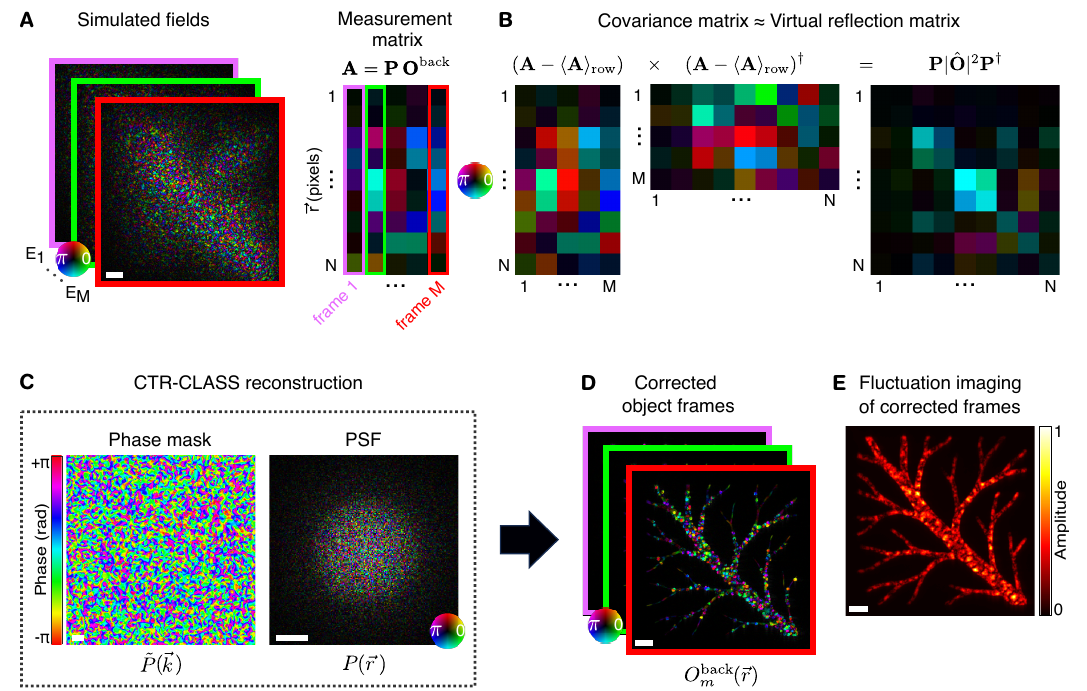}
\caption{\textbf{Reconstruction pipeline for Dynamic-target imaging through a static scattering medium with numerical results.}
\textbf{(A)} Flowing structures ($15~\mu$m diameter) in vessel-like geometry, 7~mm behind a fixed random phase mask (scattering angle $1.5^\circ$) are illuminated by a plane wave ($\lambda = 632.8$~nm). The back-reflected fields from the object propagate through the mask. Representative sampled fields at the object plane are shown, exhibiting severe aberration. Each field (color coded) is reshaped into a column of the measurement matrix $\mathbf{A} 
= \mathbf{P}\mathbf{O}^\text{back}$, where $\mathbf{P}$ is the scattering matrix and $\mathbf{O}^\text{back}$ contains the temporally varying object back-reflected fields.
\textbf{(B)} The temporal covariance matrix $(\mathbf{A}-\langle\mathbf{A}\rangle_{row})(\mathbf{A}-\langle\mathbf{A}\rangle_{row})^{\dagger}$, 
where $\langle\mathbf{A}\rangle_{row}$ denotes row (temporal per-pixel) mean, yields 
$\mathbf{P}|\hat{\mathbf{O}}|^{2}\mathbf{P}^{\dagger}$. $\hat{\mathbf{O}}$ represents object back-reflection fluctuations, after temporal mean subtraction. 
Under spatially uncorrelated dynamics, $|\hat{\mathbf{O}}|^{2}$ becomes nearly diagonal, making the covariance matrix approximately equal to a virtual reflection matrix suitable for CLASS reconstruction.
\textbf{(C)} The scattering phase mask and  PSF are reconstructed using CTR-CLASS algorithm.
\textbf{(D)} Raw frames are deconvolved with the reconstructed PSF, yielding corrected dynamic object frames.
\textbf{(E)} Temporal standard deviation image of the corrected frames highlights the 
outline geometry, despite residual speckle noise. Scale bar: 0.1 mm.}
    \label{fig:S1_simulation}
\end{figure}

To demonstrate the equivalence established in Eq.~(\ref{eq:Rvirt}), we apply the memory-efficient implementation of the CTR-CLASS algorithm~\cite{weinberg2024fluorescence} to numerically simulated dynamic-target data, as illustrated in Fig.~\ref{fig:S1_simulation}. The simulated scene consists of $M = 180$ realizations of flowing structures ($15~\mu$m diameter) in a vessel-like geometry, illuminated by a plane wave ($\lambda = 632.8$~nm) and placed $7$~mm behind a static random phase mask ($1.5^\circ$ scattering angle). The forward and backward propagation paths were modeled by two independent phase masks to account for the spatially distinct regions of the diffuser sampled in each direction (as a modeling choice rather than a method requirement).

The $M$ captured fields, each of $300 \times 300$ pixels (pixel size $3~\mu$m, 100-pixel zero-padding on each side), are reshaped into columns and stacked to form the measurement matrix $\mathbf{A} = \mathbf{P}\mathbf{O}^{\text{back}}$ (Fig.~\ref{fig:S1_simulation}A). The fields exhibit scattering distortions that render the object unrecognizable.  The CTR-CLASS algorithm is applied to $\mathbf{A}$ to recover $\tilde{P}(\vec{k})$, the scattering phase mask in the K-space, and ${P}(\vec{r})$, the coherent  PSF (Fig.~\ref{fig:S1_simulation}C). Each raw measurement is then deconvolved with the reconstructed PSF, yielding the corrected object frames $O_m^{\text{back}}(\vec{r})$ (Fig.~\ref{fig:S1_simulation}D). A temporal standard deviation image of the corrected frames (Fig.~\ref{fig:S1_simulation}E) clearly highlights the vessel-like geometry, confirming successful reconstruction of the dynamic target.

\subsubsection*{MTF estimation for incoherent reconstruction (I-CLASS)}

The incoherent reconstruction pipeline (I-CLASS) includes an additional Wiener deconvolution step (Eq.~\ref{deconv}). Following the procedure described in Ref.~\cite{weinberg2024fluorescence}, we estimate the MTF directly from the temporal fluctuations of the measurement matrix. 
Specifically, defining the fluctuation matrix $\hat{\mathbf{A}} = \mathbf{A} - \langle\mathbf{A}\rangle_{\text{row}}$, 
where each column $m$ ($m = 1, \ldots, M$) corresponds to a single recorded frame $\hat{\mathbf{A}}_{:,m}$, the MTF is estimated as:

\begin{equation}
\mathrm{MTF}(\vec{k})
\propto
\sqrt{
\sum_{m=1}^{M}
\left|
\mathcal{F}
\left\{
\hat{\mathbf{A}}_{:,m}
\right\}
\right|^2
}.
\end{equation}

\noindent
where $\mathcal{F}\{\cdot\}$ denotes the Fourier transform, and the term under the square root corresponds to the accumulated power spectrum of the temporal fluctuations across all recorded frames. 

\subsection*{S3: Model-based gradient-ascent optimization reconstruction process}

Based on the equivalence established in the Principle section, our reconstruction process can also follow a computational wavefront shaping framework \cite{Haim2025}. This approach consists of three main stages: (a) proposing a correction phase mask, (b) computing an image quality metric, and (c) iteratively updating the phase mask via gradient ascent to maximize image quality. Unlike the CTR-CLASS algorithm, this method directly optimizes a digital "virtual SLM" correction pattern without the need to compute covariance matrices.

Given the sequence of $M$ temporal field measurements $E_m(\vec{r})$ (where $m = 1, \ldots, M$), at each iteration $t$, a correction phase mask $\phi_t(\vec{r})$ is applied to compensate for the unknown scattering (assuming $E_m(\vec{r})$ represents the field at the diffuser plane; otherwise, the field should be propagated to this plane beforehand). Each measured field is back-propagated through this correction mask to reconstruct the object field:
\begin{equation}
O'_{m,t}(\vec{r}) = \mathcal{P}_{-z_\text{prop}}\{E_m(\vec{r}) \times e^{-i\phi_t(\vec{r})}\},
\end{equation}
\noindent
where $O'_{m,t}(\vec{r})$ is the reconstructed object field for the $m$-th temporal frame at iteration $t$, $\phi_t(\vec{r})$ is the correction phase mask at iteration $t$, $\mathcal{P}_{-z_\text{prop}}\{\cdot\}$ represents the propagation operator for back-propagation over distance $z_\text{prop}$ from the diffuser plane to the object plane \cite{goodman2005introduction}. An incoherently compounded image is then formed by averaging the intensity of all corrected fields:
\begin{equation}
I_t(\vec{r}) = \left\langle |O'_{m,t}(\vec{r})|^2 \right\rangle_m = \frac{1}{M}\sum_{m=1}^{M} |O'_{m,t}(\vec{r})|^2,
\end{equation}
\noindent
where $I_t(\vec{r})$ is the incoherently compounded intensity image at iteration $t$, and $\langle \cdot \rangle_m$ denotes temporal averaging over all $M$ measurements.
The quality of $I_t(\vec{r})$ is quantified using a metric $\mathcal{Q}_t$ that calculates the image variance (contrast), known to improve with proper scattering compensation. Other metrics such as entropy and Fourier-domain variance can also be used. The correction phase is then updated via gradient ascent to maximize this metric:
\begin{equation}
\phi_{t+1}(\vec{r}) \leftarrow \phi_t(\vec{r}) + \alpha \frac{\partial \mathcal{Q}_t}{\partial \phi_t(\vec{r})},
\end{equation}
\noindent
where $\alpha$ is the step size (learning rate), $\frac{\partial \mathcal{Q}_t}{\partial \phi_t(\vec{r})}$ is the gradient of the quality metric with respect to the phase mask, computed efficiently using automatic differentiation through the entire computational graph from $\phi_t(\vec{r})$ to $\mathcal{Q}_t$. This process iterates until convergence or is stopped at a predefined maximum number of iterations, yielding both the optimized phase correction and the sequence of corrected dynamic object frames.

While both the matrix-based and model-based approaches exploit the same mathematical equivalence, they differ in computational strategy. The optimization framework may require fewer measurements but necessitates careful parameter tuning. Both methods can accommodate anisoplanatic scattering via multi-layer models \cite{Haim2025, kang2023tracing} in the coherent imaging regime.

\subsection*{S4: Handling quasi-static components }

When the scene contains quasi-static components, such as a slowly drifting reflectivity,
the temporal decorrelation assumption (Eq.~\ref{eq:decorr}) is violated. This is analogous to illuminating an object with correlated patterns in conventional matrix-based imaging~\cite{lee22}. In such cases, even after temporal mean subtraction, the reconstruction may fail or be degraded.

To address this, we apply differential imaging by subtracting consecutive measured fields (Eq.~\ref{eq:Em}):
\begin{equation}
\Delta E_m(\vec{r}) = E_{m+1}(\vec{r}) - E_m(\vec{r}) = \int P(\vec{r}, \vec{r}')\Delta O_m^{\text{back} }(\vec{r}') d\vec{r}',
\label{eq:diff}
\end{equation}
\noindent
where, since $P(\vec{r}, \vec{r}')$ is static in time, the subtraction acts only on the object field, with $\Delta O_m^{\text{back}}(\vec{r}') = O_{m+1}^{\text{back}}(\vec{r}') - O_m^{\text{back}}(\vec{r}')$ and the differential measurement matrix is:\\ $ \mathbf{A}_{\text{diff}} = \bigl[\Delta \mathbf{E_1},\, \Delta \mathbf{E_2},\, \ldots,\, \Delta \mathbf{E}_{M-1}\bigr] = \mathbf{P}\,\Delta\mathbf{O}^{\text{back}}$.

This acts as a high-pass temporal filter and cancels slowly varying components between consecutive frames. The differential fields therefore satisfy the decorrelation requirement of Eq.~(\ref{eq:decorr}), and the full reconstruction pipeline of S2 and S3 can be applied to $\mathbf{A}_{\text{diff}}$ without modification.

\subsection*{S5: Patch-based reconstruction for anisoplanatic imaging}

To address anisoplanatic aberrations where the memory effect range is limited, we employ a patch-based reconstruction strategy (Fig.~\ref{fig2_tube} and Fig.~\ref{fig3_finger}). The recorded complex field is first numerically propagated from the diffuser plane to the estimated object plane, where the object features are spatially localized. In this plane, we isolate specific regions of interest (ROIs) automatically (Fig.~\ref{fig2_tube}) or manually (Fig.~\ref{fig3_finger}), each corresponding to an approximate isoplanatic patch.

Since the information from a localized object patch spreads spatially due to diffraction when propagated back to the diffuser plane, the computational grid must be large enough to contain this spread. Therefore, we isolate the ROI either by zeroing out the field outside the selected patch (while maintaining the full array size) or by cropping the ROI and applying zero-padding to restore the necessary spatial support. The isolated field is then propagated back to the diffuser plane, and the correction algorithm is applied to retrieve the local phase mask.

To form the final full-field image, the independently reconstructed patches are stitched together. In regions where patches overlap, we select the pixel value with the maximum amplitude to ensure optimal contrast and seamless transitions.

It should be noted that this 2D mosaicking approach is effective when the scattering medium is relatively thin or the anisoplanatism is moderate. In regimes of strong scattering, where light from different object regions is heavily mixed and the memory effect is negligible, this planar segmentation may be insufficient. In such cases, utilizing a multi-layer phase mask model would be required to accurately disentangle the scattering contributions.


\clearpage 

\paragraph{Movie S1. Dynamic coherent imaging of flowing beads through scattering.}
This movie shows the uncorrected object-plane fields together with their corresponding reconstructions obtained using the matrix-based (CLASS) and model-based computational wavefront-shaping methods for the experiment presented in Fig.~\ref{fig2_tube}.

\paragraph{Movie S2. I-CLASS reconstruction of the reduced fluorescence flowing-beads dataset.}
This movie shows the fluorescence imaging experiment presented in Fig.~\ref{fig4_incoherent}. The measured frames acquired through the scattering layer are displayed alongside the corresponding I-CLASS reconstructions obtained from the reduced data set ($M=198$ realizations) after excluding frames containing isolated beads that became directly visible after the initial reconstruction.

\paragraph{Movie S3. Matrix-based reconstruction of the complete fluorescence flowing-beads sequence.}
This movie shows the complete fluorescence imaging sequence presented in Fig.~\ref{fig4_incoherent}. The uncorrected and reconstructed frames are displayed side by side. The reconstruction is performed using the correction estimated from the reduced dataset used for Movie~S2.

\end{document}